\documentclass[useAMS,usenatbib]{mn2e}
\usepackage{graphicx}

\title[NIR RR Lyrae distance to M5]{Distance to Galactic globulars using the near-infrared magnitudes 
of RR Lyrae stars: IV. The case of M5 (NGC~5904)}
\author[G. Coppola et al.]{G. Coppola$^{1}$\thanks{E-mail: coppola@na.astro.it}, M. Dall'Ora$^{1}$,  V. Ripepi$^{1}$, M. Marconi$^{1}$, I. Musella$^{1}$, G. Bono$^{2,3,4}$,
\newauthor
A. M. Piersimoni$^{5}$, P. B. Stetson$^{6}$ and J. Storm$^{7}$\\
$^{1}$ INAF--OACN, Via Moiariello 16, 80131, Napoli, Italy\\
$^{2}$Dipartimento di Fisica-- Univ. di Roma Tor Vergata, via della Ricerca Scientifica 1, 
00133 Roma, Italy\\
$^{3}$ INAF--Osservatorio Astronomico di Roma, Via Frascati 33, 00040, Monte Porzio Catone, Italy\\
$^{4}$ European Southern Observatory, Karl-Schwarzschild-Str. 2, 85748 Garching bei Munchen, Germany\\
$^{5}$INAF--OACTe, via M. Maggini, 64100 Teramo, Italy\\
$^{6}$DAOHIA, NRC, 5071 West Saanich Road, Victoria, BC V9E 2E7, Canada\\
$^{7}$AIP, An der Sternwarte 16, D-14482 Potsdam, Germany\\
}

\newcommand{\pasp}{PASP}
\newcommand{\aj}{AJ}
\newcommand{\mnras}{MNRAS}
\newcommand{\apjs}{ApJS}
\newcommand{\apj}{ApJ}
\newcommand{\apjl}{ApJL}
\newcommand{\araa}{AR\&A}
\newcommand{\aap}{A\&A}

\newcommand{\aaps}{A\&AS}
\newcommand{\aapr}{A\&AR}
\def\ngc#1{NGC$\,$#1}

\begin{document}

\date{Accepted: Received:}

\pagerange{\pageref{firstpage}--\pageref{lastpage}} \pubyear{2011}

\maketitle

\label{firstpage}

\begin{abstract}
We present new and accurate near-infrared (NIR) $J$, $K$-band time series data 
for the Galactic globular cluster (GC) M5 = \ngc{5904}. Data were collected with SOFI at the NTT 
(71 $J$ $+$ 120 $K$ images) and with NICS at the TNG (25 $J$ $+$ 22 $K$ images) and cover 
two orthogonal strips across the center of the cluster of 
$\approx 5 \times 10$ arcmin$^{2}$ each. These data allowed 
us to derive accurate mean $K$-band magnitudes for $52$ fundamental (RR$_{ab}$) 
and $24$ first overtone (RR$_{c}$) RR Lyrae stars. Using this sample 
of RR Lyrae stars, we find that the slope of the $K$-band Period Luminosity 
(PLK) relation ($-2.33 \pm 0.08$) 
agrees quite well with similar estimates available in the literature. We also find, 
using both theoretical and empirical calibrations of the PLK relation, a true 
distance to M5 of $14.44 \pm 0.02$ mag. This distance modulus agrees very well  
(1$\sigma$) with distances based on main sequence 
fitting method and on kinematic method ($14.44 \pm 0.41$ mag, \citealt{rees_1996}), while is systematically smaller than the distance based on the 
white dwarf cooling sequence ($14.67 \pm 0.18$ mag, \citealt{layden2005}), even if with a difference slightly larger than 1$\sigma$.         
The true distance modulus to M5 based on the PLJ relation ($14.50 \pm 0.08$ mag) is in quite good agreement
with the distance based on the PLK relation further 
supporting the use of NIR PL relations for RR Lyrae stars to improve the precision 
of the GC distance scale. 
\end{abstract}


\section{Introduction}\label{sec_intro}
Galactic Globular Clusters (GGCs) are crucial stellar systems to constrain the
input physics adopted to construct evolutionary and pulsation
models~\citep{renzini1988,marconi2003,cassisi2010}, to investigate the kinematic
properties of gas-poor, compact systems~\citep{meylan1997} and to constrain the
formation and evolution of the Galactic spheroid (halo, thick disk, bulge;
e.g.~\citealt{mackey2005,bica2006,forbes2010}).  The GGCs often host RR Lyrae
variables (RRLs), which are fundamental distance indicators for low-mass, old
stellar populations. They are bright enough to have been detected in several
Local Group galaxies
(e.g.~\citealt{dallora2003,dallora2006,pietrzynski2008,greco2009,fiorentino2010,yang2010}).  
They can also be easily identified, since they have characteristic light curves and periods. The
most popular methods to estimate their distances are the visual
magnitude -- metallicity relation and the near-infrared (NIR) Period-Luminosity
(PL) relation (e.g.~\citealt{bono2003,cacciari2003}). The reader interested in
other independent approaches based on RRL to estimate stellar distances is referred to the
thorough investigations by, e.g.,~\citet{marconi2003,dicriscienzoetal_2004apj,feast2008}.  

The visual magnitude -- metallicity approach appears to be hampered by several theoretical and empirical uncertainties affecting both the zero-point and the slope~\citep{bonoetal_2003mnras}. The NIR PL relation seems very promising, since it shows several indisputable advantages. Dating back to the seminal investigation by~\citet{longmoreetal_1986mnras} it has been demonstrated on empirical basis that RRL do obey to a well defined $K$-band PL relation. The key reason why the PL relation shows up in the NIR bands is mainly due to the fact that the bolometric correction in the NIR bands, in contrast with optical bands, steadily decreases when moving from hotter to cooler 
RRLs. This means that they become brighter as they get cooler. The pulsation periods --at fixed stellar mass and luminosity-- become longer, since cooler RRLs have larger radii. The consequence is a stronger correlation between period and magnitude when moving from the $I$- to the $K$-band.        

The advantages of using NIR PL relations to estimate distances are manifold.\\
{\em i)} Evolutionary and pulsation predictions indicate that the NIR PL 
relations are minimally affected by evolutionary effects inside the RRL instability strip. The same outcome applies for the typical spread in mass inside the RRL instability strip~\citep{bonoetal_2001mnras, bonoetal_2003mnras}. This means that individual RRL distances based on the NIR PL relations are minimally affected by systematics introduced by their intrinsic parameters and evolutionary status.  
{\em ii)} Theory and observations indicate that fundamental (F or RR$_{ab}$) and first overtone (FO or RR$_{c}$) RRL do obey independent NIR PL relations that are linear over the entire period range covered by F and FO pulsators.  

The NIR PL relations together with the aforementioned features have also three positive observational advantages. \\
{\em a)} The NIR magnitudes are minimally affected by reddening uncertainties. This means that the RR Lyrae NIR PL relations can provide robust distance estimates for systems affected by differential reddening.   
{\em b)} The luminosity amplitude in the NIR bands is at least a factor of 2-3 smaller than in the optical bands. Therefore, accurate estimates of the mean NIR magnitudes can be obtained with a modest number of observations. Moreover, empirical light curve templates~\citep{jonesetal_1996pasp} can be adopted to improve the accuracy of the mean magnitude even when only a single observation is available. 
{\em c)} Thanks to the unprecedented effort by the 2MASS project~\citep{skrutskie_2006AJ}, accurate samples of local NIR standard stars are available across the sky. This means that both relative and absolute NIR photometric calibrations do not require supplementary telescope time.    

The use of the NIR PL relations is also affected by four drawbacks. \\  
{\em --} Empirical estimates of the slope of NIR PL relations show a significant scatter from cluster to cluster. They range from $\sim -1.7$ (IC$4499$~Sollima et al.~\citeyear{sollimaetal_2006mnras}) to $\sim -2.9$ (M$55$,~Sollima et al.~\citeyear{sollimaetal_2006mnras}) and it is not clear yet whether this change is either intrinsic or caused by possible observational biases. 
{\em --} Both theoretical and empirical investigations of the PLK show a not-negligible scatter of the zero point. This is a crucial point, as we plan to adopt the PLK as a tool to derive distances. Indeed, if we set as a reference $logP=-0.5$ and $[Fe/H]=-1.5$, the most robust absolute magnitude estimates range from $M_{K,-0,5,-1,5}= -0.33$ (infrared flux method, \citealt{longmoreetal_1990mnras}) to $M_{K,-0,5,-1,5}= -0.46$ (HB models, \citealt{catelanetal_2004ApJS}; empirical calibration \citealt{sollimaetal_2006mnras}), with intermediate values of $M_{K,-0,5,-1,5}= -0.39$ (pulsational models \citealt{bonoetal_2001mnras}; HB models,\citealt{cassisietal_2004AA}).\\
{\em --} Evolutionary and pulsation predictions indicate that the intrinsic spread of the NIR PL relations decreases  as soon as either the metallicity or the HB-type of the Horizontal Branch (HB) is taken into account~\citep{bonoetal_2003mnras, cassisietal_2004AA, catelanetal_2004ApJS, delprincipeetal_2006ApJ}. This means that accurate distance estimates of field RRLs do require an estimate of the metallicity. Moreover, no general consensus has been reached yet concerning the value of the coefficient of the metallicity term in the NIR Period-Luminosity-Metallicity (PLZ) relations. The current estimates for the $K$-band range from $0.08$~\citep{sollimaetal_2006mnras} to $0.23$~\citep{bonoetal_2003mnras} 
mag/dex based on cluster and field RRLs, respectively. \\
{\em --} The use of the template light curves does require for each object accurate estimates of B/V-band amplitudes and of the epoch of maximum.

To address these problems our group undertook a long-term project aimed at providing homogeneous and accurate NIR photometry for several GCs hosting sizable samples of RRLs and covering a wide range of metallicities. We have already investigated the old LMC cluster Reticulum~\citep[][hereinafter paper I]{dalloraetal_2004ApJ}, the GGC M92~\citep[][hereinafter paper II]{delprincipeetal_2005AJ} and $\omega$ Centauri~\citep[][hereinafter paper III]{delprincipeetal_2006ApJ}.  

In this paper we present new results for the GGC M5 (NGC$5904$). This system is a typical \textit{halo} globular cluster, and indeed, it is located at a distance of only $\sim 7 \, \rm kpc$ from the Sun, $5.5 \, \rm kpc$ from the Galactic Center, and $4.9 \, \rm kpc$ above the Galactic disc~\citep{zinn_1985ApJ}; its space motion is $\sim 490 \, \rm km \rm s^{-1}$ \citep{cudworthetal_1993AJ}. Moreover, it has a low reddening ($E(B-V)=0.03$, according to the Harris catalog, ~\citealt{harris_1996aj}, and its new revision~\citealt{harris_2010aj}) and metal-intermediate composition, with estimates ranging from $-1.0 \, \rm dex$~\citep{butler_1975ApJ} to $-1.346 \, \rm dex$~\citep{carrettaetal_2009A&A}. This cluster is a very good target to investigate the NIR PLZ relation, since it hosts a rich sample of RRLs  ($\sim 130$ in the 2002 release of the Clement's on-line catalog,~\citealt{clement}). The RR$_{ab}$ stars have an average period of $\sim 0.55 \ \rm d$, making M5 as one of the classical Oosterhoff type I clusters~\citep{oosterhoff_1939Obs}. It is located only two degrees from the celestial equator, and therefore, the search for cluster variables 
has been performed using telescopes from both hemispheres. According to~\citet{hogg_1973PDDO} a total of $97$ variables were present in M5 and among them $93$ were RRLs. A significant fraction of these variables were discovered by~\citet{bailey_1917AnHar} with the remaining stars identified by~\citet{1941AnLei..17D...1O}. The central regions of M5 have been surveyed 
photographically by~\citet{gerashchenko_1987IBVS} and~\citet{kravtsov_1988ATsir} who found an additional $27$ variables. Subsequently,~\citet{cohenetal_1992PASP} supplemented the catalog with five other variables, while~\citet{reid_1996MNRAS} presented observations of $49$ RRLs, three of which were new discoveries. \citet{brocatoetal_1996AJ} added $15$ new variables located across the central regions of the cluster. More recent investigations by~\citet{sandquistetal_1996ApJ} 
and~\citet{drisse&shara_1998AJ} identified $28$ previously unknown variables, while~\citet{kaluznyetal_2000A&AS} and~\citet{caputoetal_1999MNRAS} identified $32$ new
variables. 
We end up with a sample of $102$ variable stars, and among them $71$ are F and $31$ FO RRLs. 

The paper is organized as follows: in Sec.~\ref{sec_obs} we discuss the observations and the strategy adopted to perform the photometry and to calibrate the data. The RRL properties and the approach adopted to determine their light curves are presented in Sec.~\ref{sec_rrl},
while in Sec.~\ref{sec_pl} we present the PLK and the PLJ relations. 
Finally, in Sec.~\ref{sec_con}, we summarize the current findings and briefly outline future perspectives.

\section{Observations and data reduction}\label{sec_obs}

\begin{figure}
\includegraphics[width=8cm,height=8cm]{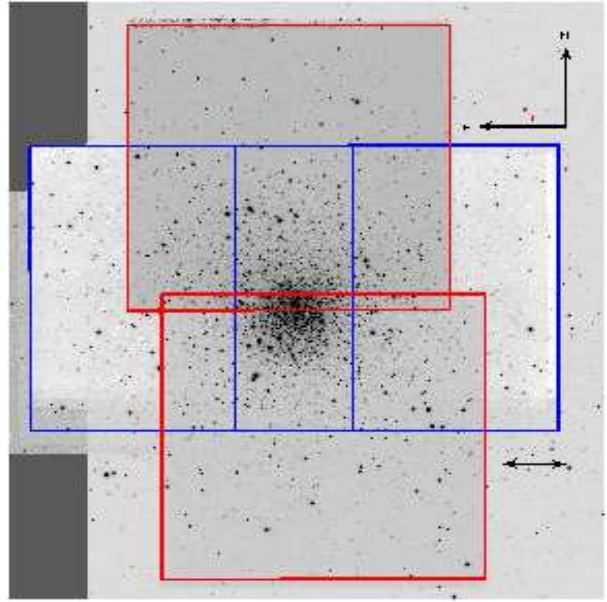}
\caption{The coverage of the two data sets collected with the SOFI@NTT/ESO telescope (red squares)
and NICS@TNG (blue squares), superimposed to M5. The showed picture has been obtained by aligning the NIR images to a WFI@2.2 MPG/ESO \textit{I-}band reference frame}\label{fig_map}
\end{figure}

We collected $J$ and $K_{s}$ images in six different runs from February 2001 to March 2002 with SOFI@NTT/ESO\footnote{http://www.eso.org/sci/facilities/lasilla/instruments/sofi/\\overview.html} and $J$ and $K'$ data in two runs in May 2005 and July 2005 with NICS@TNG\footnote{http://www.tng.iac.es/instruments/nics/}. The SOFI data were collected in the large field mode with a pixel scale of $0.292$ arcsec/pixels and with a field of view of $4'.94 \times 4'.94$. NICS was also used in the large field mode, with  pixel scale of $0.25$ arcsec/pixels and with a field of view of $4'.2 \times 4'.2$. For both instruments, observations were obtained of off-target fields for the sky subtraction. 
Since the cluster is moderately extended (tidal radius $\sim 28.4$ arcmin,~\citealt{harris_2010aj}), for both instruments two different pointings were observed, mapping the cluster in the North-South direction (SOFI) and in the East-West direction (NICS), covering two strips of about $5$ $\times$ $5$ $\rm arcmin^{2}$ in both directions (see Fig.~\ref{fig_map}). We collected $71$ ($228$ s of total exposure) $J$- and $120$ ($1529$ s) $K_{s}$-bands useful frames with SOFI, and $25$ ($193$ s) $J$- and $22$ ($184$ s) $K'$-bands images with NICS. The number of epochs for recovered RRLs ranges from $1$ to $16$. The log of observations is given in Table~\ref{tab_log}.

The raw SOFI frames were first corrected for the cross-talk effect with the IRAF\footnote{IRAF is distributed by the National Optical Astronomical Observatory, which is operated by the Association of Universities for Research in Astronomy, Inc., under cooperative agreement with the National Science Foundation.} procedure \texttt{crosstalk.cl}, available in the SOFI web pages. Data were then pre-processed with our IRAF-based custom pipeline, which corrects for bias, flat field and bad pixel mask, and subtracts the sky contribution with a two-step technique as described in~\citet{pietrzynskietal_2002AJ}.
The NICS images were corrected for cross-talk with a \texttt{FORTRAN} program made available in the NICS web pages, and thereafter pre-processed with the same pipeline described above, without applying any bad pixel mask.

\begin{table*}
\begin{minipage}{150mm}
\caption{Observing Log.The first column shows the date of observation, written as day/month/year. The second, third, fourth and fifth column list the instrument used, the filter, the number of epochs collected and the total exposure time, respectively. The last column reports the average seeing, computed as the median seeing measured on the individual images.}\label{tab_log}
\begin{tabular}{|c|c|c|c|c|c|}\hline 
DATE	& INSTRUMENT &	FILTER &	EPOCHS	&  TOTAL EXP. TIME &	$<$SEEING$>$ (arcsec)\\
 	&            &	       &		&      (sec)       &	  (arcsec)\\
\hline
050201	& SOFI	&	$J$ &	2 &	 18 &		0.71\\
	 &	&	$K_s$&	2	&120	&	0.74\\

070201&	SOFI&		$J$&	2&	 18	&	0.74\\
      &		&	$K_s$&	2	&108	&	0.63\\

240202&	SOFI	&	$J$&	4&	 36&		0.96\\
      &		&	$K_s$&	4	&228	&	0.75\\

250202	&SOFI		&$J$	&6	& 54	&	0.66\\
       &	&	$K_s$	&6&	348	&	0.58\\

260202&	SOFI&		$J$&	5&	 45&		0.92\\
      & &		$K_s$&	5&	300&		0.93\\

090302&	SOFI	&	$J$&	6&	 57	&	0.72\\
      &	&	$K_s$&	6&	425	&	0.71\\

130505&	NICS	&	$J$&	2&	105	&	0.50\\
      &		&	$K'$	&2	& 80&		0.45\\

180705&	NICS&		$J$&	2&	 88&		0.66\\
      &	&		$K'$&	2&	104&		0.65\\
\hline 
\end{tabular}
\end{minipage}
\end{table*}

\subsection{Photometry}\label{sec_phot}
For each image, a preliminary PSF model and a list of stars were produced with the DAOPHOTIV/ALLSTAR package~\citep{stetson_1987PASP, stetson_1994pasp}. In order to cross-match individual star catalogs, we computed geometric transformations with the DAOMATCH/DAOMASTER programs~\citep{stetson_1994pasp}, using as a reference a 2.2@MPG/ESO \textit{I-}band image, available in the ESO data archive. A master star list was therefore produced on the total stacked image. We then performed a first ALLFRAME \citep{stetson_1994pasp} run, and a second star list was built on the median of the star-subtracted input images, which was merged with the previous one, in order to get a more complete input catalog. New accurate, spatially varying, PSF models were subsequently computed on the individual images, after cleaning each PSF star of contamination by faint neighbors, and a second ALLFRAME run was performed. We ended up with a final catalog of 38,131 sources.

\subsection{Calibration}\label{sec_cal}
We calibrated each individual image to the Two Micron All Sky Survey (2MASS) photometric system, by selecting clean local 2MASS standards (i.e. bright and reasonably isolated, by taking advantage of the 2MASS photometric flags). The SOFI images collected before April 2003 are affected, due to the misalignment of the Large Field objective, by a strong distortion on a strip of $\sim 200$ pixels wide along the $x$-axis. We found that the difference between instrumental and 2MASS magnitudes can attain values of the $\sim 0.3$ magnitudes in this region of the detector. Therefore, we wrote an ESO-MIDAS (Munich Image Data Analysis System) procedure to simultaneously correct for this positional 
effect and to get standard photometric zero-points. The procedure performs a polynomial regression of the instrumental magnitude as a function of the $x,y$ and 2MASS magnitude variables. The accuracy of the calibration, evaluated as the standard deviation of the fit, varies with the images from approximately $0.01$ to $0.06$ mag. We explicitly note that a spatially 
varying PSF might introduce a positional effect. However, to overcome this problem the 
selected PSF stars for each image are uniformly distributed across the frame. The difference 
between instrumental and 2MASS magnitudes in the region not affected by distortions is 
minimal. Therefore the positional effect seems to be caused by the objective misalignment. 

We did not include color terms in the calibration, since they are negligible in the 
range of colors of interest. The SOFI data have minimal color dependence, as shown in the 
SOFI web 
pages\footnote{http://www.eso.org/sci/facilities/lasilla/instruments/sofi/\\inst/setup/Zero\_Point.html}, 
while the NICS $K'$-band data typically have a non-negligible dependence on the color, which in the original transformation by \citealt{wainscoat92} is given in terms of the $H-K$ color. In fact, by combining the transformation between 
the $K'$ and the standard $K$ filter in the Caltech photometric system 
(CIT,~\citealt{frogel1978}), as described in~\citet{wainscoat92}, with the transformations 
between the 2MASS and the CIT systems~\citep{carpenter2001}, we end up with the following \
transformation: $K_{2M}=K'-0.2 \times (H-K)_{CIT}$. 
Since the typical $(H-K)$ HB colors range from $-0.1$ to $+0.05$, and considering that the RRL Instability Strip is confined to an even narrower color range, we conclude that the maximum systematic error is $\sim 0.01$ magnitudes.

\section{The Color-Magnitude Diagram}
\begin{figure*}
\includegraphics[width=9cm,height=9cm]{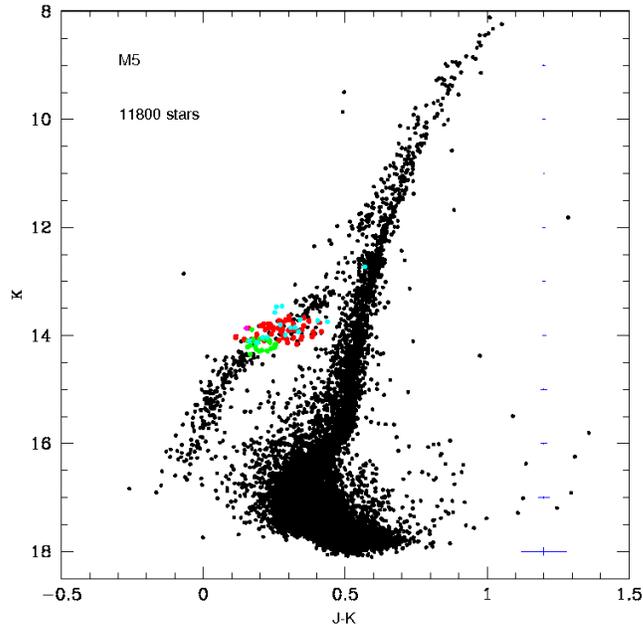}
\caption{$K$, ($J-K$  Color-Magnitude Diagrams of the globular M5. 
Red and green circles mark the position of RR$_{ab}$ and RR$_{c}$ variables, respectively. 
The cyan filled circles show variables neglected in the estimate of the distance. 
The error bars on the right display intrinsic errors in magnitude and in color. 
The number of selected stars is also labeled.}\label{fig_cmd}
\end{figure*}

Fig.~\ref{fig_cmd} shows the observed $K$-($J-K$) color-magnitude diagram. 
Stars plotted in this figure were selected by choosing only stars with small photometric 
contamination by close companions, using the \texttt{SEPARATION} index~\citep{stetsonetal_2003pasp}. 
The contamination limit was set to $3$, meaning that we selected only stars whose ratio (expressed 
in magnitudes) between the central surface brightness and the sum of the brightnesses of all other 
stars out to $10 \times$ the FWHM is equal or larger than $3$ ($=$ a factor of $16$ in flux). This means that we selected only 
stars whose correction for photometric contamination by other stars was smaller than $\sim 15\%$. 
Since ALLFRAME fits the PSF model on each individual star when all the other stars in the image 
have been digitally subtracted, any uncorrected photometric contamination remaining should therefore be much smaller. 
We show only stars with estimated ALLFRAME photometric standard errors smaller than $0.04$ mag, both in the $J$ and $K$ band, corresponding to a signal-to-noise ratio on individual images of $S/N \sim 5$. Below this limit, we consider that the measurements on the individual images do not have sufficient accuracy to be 
included in our catalog. Adopting these cuts, we end up with $11800$ star-like sources. Red and green filled circles represent RR$_{ab}$ and RR$_{c}$ stars 
(see below), while cyan filled circles show variable stars excluded from the PLK and PLJ 
analysis (see Sec.~\ref{sec_pl}). Our photometry covers the full range of magnitudes running from the Tip of the Red 
Giant Branch (RGB, $K \sim 8$) down to $\approx 0.5$ magnitude below the Turn-Off (TO, 
$K\sim 17$) region. The photometric limit can therefore be approximately located at 
$K \approx 18$ mag, where the intrinsic photometric uncertainty is of the order of 
$\sim 0.03$ mag in the $K$ band. The CMD clearly shows the \textit{bump} 
of the RGB, at $K \approx 13$ mag, as well as the separation between the RGB and the 
Asymptotic Giant Branch ($K \approx 12.5$). The intrinsic features of the CMD and a
comparison with evolutionary predictions will be addressed in a forthcoming paper.

\section{RR Lyrae stars}\label{sec_rrl}
We recovered $102$ out of $131$ RR Lyrae variables on the basis of the WCS positions 
listed in the catalogs by~\citet{evstigneevaetal_1999yCat} and~\citet{samusetal_2010yCat}. 
The $29$ RRLs not included in the current sample are located outside the region covered 
by our data. Barycentric Julian days were computed for each epoch of observation. 
The total exposure for each epoch was split into a number of shorter frames, and the 
final mean magnitude was estimated as an average of the measurements over all 
the individual frames. The $K$-band light curves of RRLs are almost 
sinusoidal and the luminosity amplitude in this band is also modest. However, 
the most accurate results for the mean magnitudes are achieved by using 
the light curve templates provided by \cite{jonesetal_1996pasp}. The use of the
template  yields accurate mean magnitudes even when only a single epoch is available. 
Note that the use of the template requires accurate ephemerides and 
$B$-band amplitudes. Therefore, we mined the available literature to get 
updated periods, epochs of maxima and $B$-band amplitudes. In most cases 
the $B$-band amplitude was not available, and we transformed the $V$-band 
amplitude into the $B$-band one using the relation $A_{B}=1.264*A_{V} +0.028$ 
\citep{jonesetal_1996pasp}. Table~\ref{tab_rrl} lists our adopted parameters: 
the name of the variable (following the number scheme proposed in~Caputo et al.~\citeyear{caputoetal_1999MNRAS}), 
the adopted period with the related reference, the epoch of maximum with the associated reference, the computed 
intensity-averaged $K$-band magnitude with the uncertainty $\sigma$, the variable 
type (fundamental, first overtone or Blazkho RRLs) and the $B$-band 
amplitude $A_{B}$, with the reference. For the Blazkho RRLs we follow the classification 
proposed by~\cite{jurcsik_2010arXiv1010}.  Moreover, in several cases the period and the 
epoch of maximum were not available from the same reference, and we imposed a shift to the phased light curve to 
reasonably match the template. Shifts are listed in column $8$ of Table~\ref{tab_rrl}. Finally, column $9$ lists the adopted $V$-band magnitude as extracted by the optical light curves of the unpublished archive of author PBS.
In some cases the epoch was not available in literature and we used the epoch of 
our first measure to calculate the intensity-averaged $K$-band magnitude. It is worth noting that period changes among the M5 RRLs have been 
detected by ~\cite{stormetal_1991PASP,reid_1996MNRAS,szeidl_2010arXiv1010}, but these changes 
are minimal and do not affect the conclusions of this investigation. An updated and homogeneous 
photometric catalog of the M5 RRLs is highly desirable.  Note that RRLs for which the $B/V$ amplitude 
was not available in the literature were neglected. The intensity-weighted mean magnitudes --computed 
by DAOMASTER-- of these objects are listed in column $4$ of Table~\ref{tab_rrl}. We also excluded 
a few variables that were heavily contaminated by bright neighbors or by close companions 
(see Sec.~\ref{sec_noterrl}). The distance estimates based on the PLK relation rely on 
$52$ RR$_{ab}$ and $24$ RR$_{c}$ variables. Some example of RRL light curves are depicted in 
Fig.~\ref{fig_curve}. Black and red filled circles represent the SOFI and the NICS observations, 
respectively. The complete atlas of the light curves is available in the electronic version of this paper.

\begin{table*}
\begin{minipage}{150mm}
\caption{Summary of the light curve parameters and properties of the M5 RRLs. \textit{Col.} $1$: identification star;
\textit{Col.} $2$: adopted period with reference; \textit{Col.} $3$: adopted epoch with reference;
\textit{Col.} $4$:  derived intensity-averaged $K$-band magnitude; \textit{Col.} $5$: derived uncertainty on intensity-averaged $K$-band magnitude;  \textit{Col.} $6$: type; \textit{Col.} $7$: adopted $B$-band amplitude with reference; \textit{Col.} $8$: adopted shift of the phase; \textit{Col.} $9$: adopted $V$-band magnitude as extracted by light curves of the private archive of author PBS.}\label{tab_rrl}
\begin{tabular}{|l|l|l|l|c|l|l|c|c|l|}\hline
Var & Period (Ref.)  & Epoch (Ref.)   & $<K>$ & $\sigma$ & Type &  A$_{B}$ (Ref.) & $\Delta \phi$ & (V) & Note \\
 & [d] & (JD) & mag & mag &  &  &  &  &\\
      \hline 
V1   &   0.5217868  (S10) &  48399.795   (R96) &   13.981  &  0.003   & Bl    &  1.43 (K00) & -0.05 & 15.238& no PLJ\\
V4   &   0.449699   (K00) &  48399.732   (R96) &   14.169  &  0.003 &   Bl    & 1.24 (K00) & -0.40 & 15.043&\\
V5   &   0.54584    (R96) &  48399.787   (R96) &   13.920  &  0.003  &   Bl    & 1.42 (R96)& -0.05&  15.188&no PLJ \\
V6   &   0.54882     (R96) &   48399.743     (R96) & 13.853   &   0.002  &   RR$_{ab}$  & 1.18 (R96) & -0.30&15.225     \\
V7   &  0.494404    (S91) &   46931.867  (S91) & 14.031  &  0.003 &   RR$_{ab}$   &1.58 (S91)  & +0.10    	&  14.959        &no PLJ\\
V11  &  0.595911    (K00) &  48399.800   (R96) & 13.714  &  0.009 &   RR$_{ab}$   &1.43 (K00)  & +0.12    	&  14.949           &no PLJ\\
V12  &  0.46770          (R96)  &  48399.827   (R96) &  14.064 &  0.008 &   RR$_{ab}$   &1.53 (R96)  &  	      	&  15.066          &no PLJ\\
V13  &    0.51313       (R96) &    48424.744     (R96) &    13.848    &    0.003   &   RR$_{ab}$  &  1.40   (R96)& -0.40&  14.861      &\\
V14  &    0.48733       (R96) &    48399.764     (R96) &    14.036    &    0.004   &   Bl    &  1.64   (R96) &-0.05  &   15.119 &\\
V16  &  0.647634    (K00) &  48400.734   (R96) & 13.730  &  0.003 &   RR$_{ab}$   &1.53 (K00)  &	      	&  14.887          &no PLJ\\
V17  &  0.601395    (S10) &  52331.772* & 13.793  &  0.003 &   RR$_{ab}$   &1.44 (S10)  &	 -0.20     	&  14.912         &\\
V18  &  0.464       (S91)   & 46931.722   (S91)& 13.984 &      0.010&       Bl     &  1.58 (K00) & & 15.221  &no PLJ\\
V24  &     0.48125        (R96) &     48400.702      (R96) &     14.139     &     0.003    &   Bl    &  0.96    (R96)& -0.10&   14.983 &no PLJ\\
V25  &     0.5074854      (S10) &     51947.026* &   13.940        &         &        RR$_{ab}$  &  & &   15.133 &\\
V26  &     0.623978     (S10) &     52331.892* &      13.71**    &         &        RR$_{ab}$  &  & &   14.973 &no PLK/PLJ\\
V27  &     0.47034        (R96) &     48404.890      (R96) &     14.074     &     0.002    &   Bl    &   1.44    (R96)& &  14.994 &\\
V28  &  0.543926    (S91) &  48400.710   (R96) & 13.925  &  0.012 &   RR$_{ab}$   &1.25 (K00)  &		&  15.075         &no PLJ\\
V30  &     0.592207       (K00) &     46931.320      (S91) &   13.893     &     0.003    &   Bl  &   1.07    (K00) & -0.05&    15.082  & no PLJ\\
V33  &  0.501575    (K00) &  48812.701   (R96) & 14.064  &  0.003 &   RR$_{ab}$   &1.51 (K00)  & 		&  14.917         &\\
V34  &  0.568119    (K00) &  48399.838   (R96) & 13.932  &  0.003 &   RR$_{ab}$   &1.04 (K00)  & +0.30	&  15.086   &   no PLJ    \\
V35  &  0.3081343   (S10) &  48400.745   (R96) & 14.131  &  0.003 &   RR$_{c}$    &0.61 (K00)  &	 	&  15.009          &\\
V36  &     0.626980       (R96) &     48400.818      (R96) & 13.746     &     0.003    &   RR$_{ab}$  &   0.81    (R96)& +0.20&  15.041 & \\
V37  &     0.4887954      (S10) &     51946.869* &   13.94**     &       &  RR$_{ab}$  &  & &   15.049    & no PLK/PLJ \\	
V38  &     0.4704285      (S10) &     48399.762      (R96) & 14.021     &     0.003    &   RR$_{ab}$  &  1.24    (R96)&+0.10&   15.038 &\\
V39  &  0.589035    (K00) &   51946.751*   & 13.845  &  0.003 &   RR$_{ab}$   &1.51 (K00)  &	        &  15.116  &        \\
V40  &  0.317334    (K00) &  48424.744   (R96) & 14.157  &  0.003 &   RR$_{c}$    &0.57 (K00)  &		&  15.067         &\\
V41  &  0.488577    (K00) &  52331.697* & 13.995  &  0.003 &   RR$_{ab}$   &1.37 (K00)  &     	&  15.239     &  \\
V43  &  0.66023      (R96) &     48424.762    (R96) &    13.648     &     0.026 &  RR$_{ab}$   &  0.89    (R96)& &   15.093 &no PLJ\\ 
V44  &  0.329576    (K00) &  48401.865   (R96) &  14.043  &  0.003 &   RR$_{c}$    &0.56 (K00) & 		&  14.959    &  \\ 
V45  &  0.616595    (B96) &  48400.888   (R96) &  13.627  &  0.002 &   RR$_{ab}$   &1.78 (B96) &		&  15.035       &\\
V47  &  0.539739    (K00) &  48404.910   (R96) &  13.939  &  0.003 &   RR$_{ab}$   &1.34 (K00) &		         &  15.121 & \\
V52  &     0.501785       (K00) &     48424.740      (R96) & 13.046     &     0.003    &   Bl  &  0.74    (R96)&  &     14.997          & \\
V53  &  0.373594    (B96) &    51946.832*  &  13.889  &  0.003 &   RR$_{c}$    &0.54 (B96) &		         &  14.773  &   \\
V54  &  0.454239    (K00) &  48399.732   (R96) &  14.099  &  0.003 &   RR$_{ab}$   &1.53 (K00) & +0.25	     & 15.272    &   \\
V55  &  0.3289023   (S91) &  46931.498 (S91)     &  14.113  &  0.003 &   RR$_{c}$    &0.55 (S10) & 	    &  15.101      &    \\
 V56  &     0.53468        (R96) &     48424.771      (R96) &  13.894     &     0.002    &   Bl    &  0.96    (R96)& -0.06&   15.217  &\\
V57  &  0.284673    (K00) &  48399.770   (R96) &  14.289  &  0.003 &   RR$_{c}$    &0.65 (K00) &		         &  14.995       \\
V59  &  0.542027    (K00) &  48399.740   (R96) &  13.908  &  0.004 &   RR$_{ab}$   &1.28 (K00) & 0.15	     &  15.031  &   \\
V60  &  0.285274    (O99) &   51946.726*   &  14.281  &  0.003 &   RR$_{c}$    &0.71 (S10) & 	    &   15.027   &             \\
V64  &  0.544492    (K00) &  51946.923*    &  14.050  &  0.003 &   RR$_{ab}$   &1.32 (K00) & 	    &  15.057   &       \\
V65  &     0.480758       (K00) &     48404.897      (R96) &  14.074     &     0.003    &   Bl  &    0.99    (R96)& -0.25&    15.317  &no PLJ\\
V77  &  0.8451232   (K00) &   51946.954*   &  13.450  &  0.011 &   RR$_{ab}$   &0.79 (K00) &		         &  14.706& no PLJ\\
V78  &  0.264798    (K00) &  48399.865   (R96) &  14.350  &  0.003 &   RR$_{c}$    &0.52 (K00) &		         &  15.098  &       \\
V79  &  0.333089    (K00) &  48400.725   (R96) &  14.070  &  0.003 &   RR$_{c}$    &0.47 (K00) &		         &  15.071  &       \\
V80  &  0.336549    (K00) &  48400.702   (R96) &  14.073  &  0.003 &   RR$_{c}$    &0.52 (K00) &		         &  15.064    &  \\
V81  &  0.55731406  (B96) &  48399.764   (R96) &  13.893  &  0.002 &   RR$_{ab}$   &1.21 (B96) & -0.30	    &  15.080   &  no PLJ \\
V82  &  0.558927    (K00) &  48399.791   (R96) &  13.866  &  0.003 &   RR$_{ab}$   &1.19 (K00) &		         &  15.055 &\\
V83  &     0.553329       (R96) &     48399.805      (R96) &   13.862     &     0.002    &  RR$_{ab}$   &  1.09    (R96)& +0.25&   15.085   &\\
 V85  &    0.5275226       (S10) &     51946.869*&        13.75**   &        &   RR$_{ab}$   & & &    14.861          &    no PLK/PLJ   \\	
V86  &  0.56728     (B96) &  51947.124* &  13.717  &  0.003 &   RR$_{ab}$   &1.57 (B96) & 	    &  14.731  &    no PLK/PLJ-crowded  \\
V87  &  0.7383982   (S10) &  48400.841   (R96) &  13.651  &  0.003 &   RR$_{ab}$   &0.49 (S10) &		         &  14.922 &\\
V88  &  0.328070    (K00) &  48404.910   (R96) &  14.092  &  0.003 &   RR$_{c}$    &0.57 (K00) & +0.25	    &  14.940           &\\
V89  &  0.558454    (K00) &  48400.793   (R96) &  13.852  &  0.003 &   RR$_{ab}$   &1.22 (K00) & +0.05	    &  15.169           &\\
V90  &    0.5571570       (S10) &     51946.757* &   13.86**     &        &   RR$_{ab}$   &  & &  15.060          & no PLK/PLJ  \\

\hline 
\end{tabular}
\end{minipage}
\end{table*}
\begin{table*}
\begin{minipage}{150mm}
\contcaption{}
\begin{tabular}{|l|l|l|l|c|l|l|c|c|l|}\hline
Var & Period (Ref.)  & Epoch (Ref.)   & $<K>$ & $\sigma$ & Type &  A$_{B}$ (Ref.) & $\Delta \phi$ & (V) & Note \\
 & [d] & (JD) & mag & mag &  &  &  &  &\\
      \hline 
 V91  &     0.60139      (R96) &     48401.834      (R96) &   13.787     &     0.003    &  RR$_{ab}$    &   1.44    (R96)&-0.05&  15.125  &\\
 V92  &    0.4633870       (S10) &   51946.684* &   13.99**    &       &  RR$_{ab}$  &   & &  15.199          &  no PLK/PLJ\\
 V93  &    0.552875      (O99)   &     51946.814*   &     13.73**   &        &   RR$_{ab}$  &   & & 14.967          &   no PLK/PLJ  \\
V94  &  0.51225     (R96) &  48424.788   (R96) &  13.971  &  0.003 &   RR$_{ab}$   &1.22 (R96) &		         &  15.107 &no PLJ\\
V95  &  0.2907655   (S10) &    52331.021*   &  14.222  &  0.003 &   RR$_{c}$    &0.66 (S10) &		         &  15.037   & \\
 V96  &     0.46424      (R96) &     48399.806      (R96) &   13.915     &     0.003    &   RR$_{ab}$   &  0.86    (R96)& +0.40&   15.079       & no PLK/PLJ-crowded\\
 V97  &    0.54469    (R96)    &     48399.850      (R96) &        13.891     &     0.002    & Bl       &  0.67    (R96)& &   15.000       & no PLJ \\
V98  &  0.30641   (B96) &    51946.870*  &  14.190  &  0.003 &   RR$_{c}$    &0.68 (S10) &		         &  14.958&no PLJ\\
V99  &  0.321340    (R96) &  48401.921   (R96) &  14.120  &  0.003 &   RR$_{c}$    &0.61 (R96) &		         &  14.907  &  \\
V100  &  0.294360   (R96) &  48399.761   (R96) &  14.224  &  0.004 &   RR$_{c}$    &0.72 (R96) &		         &  15.061 &\\
V103  &  0.5667     (C99) &  47629.551   (C99) &  13.771  &  0.003 &   RR$_{ab}$   &1.03 (C99) & -0.36	    &  15.019             &\\
V104  &     0.310930      (O99) &     52331.862* &   14.044       &    0.003    &   RR$_{c}$ & 0.9     (C99)&&    15.048          &  no PLK/PLJ\\
V105  &  0.2920     (C99) &  47629.813   (C99) &  14.107  &  0.007 &   RR$_{c}$    &0.94 (C99) &		         &  14.882           & \\
 V106  &     0.5624        (C99) &     47629.311      (C99) &   13.652     &     0.007    &  RR$_{ab}$   &   1.20    (C99)& 0.05&  14.804      & no PLK/PLJ-crowded  \\
 V107  &     0.5117        (C99) &     47629.776      (C99) &      13.703     &     0.005    &   RR$_{ab}$   &   1.52    (C99) &&  14.767      &  no PLK/PLJ-crowded \\
V108  &  0.329      (C99) &  51946.902* &  13.913  &  0.004 &   RR$_{c}$    &0.55 (C99) & 	    & 14.802                    & no PLK/PLJ-blended?\\
V109  &  0.476      (C99) &  47629.854   (C99) &  14.069  &  0.003 &   RR$_{ab}$   &1.55 (C99) &		         & 15.127            & no PLJ \\
V110  &  0.598528   (O99) &   51946.779*  &  13.838  &  0.003 &   RR$_{ab}$   &0.95 (C99) & 	    &  15.218             &\\
V111   &     0.6233      (C99) &     47629.283      (C99) &     13.456     &     0.009    &   RR$_{ab}$  &    0.93    (C99)& &  15.018      & no PLK/PLJ-crowded \\
V112  &  0.5367     (C99) &  47629.546   (C99) &  13.858  &  0.003 &   RR$_{ab}$   &1.29 (C99) &		         &  14.941 &\\
V113  &  0.2843     (C99) &  47629.758   (C99) &  14.185  &  0.003 &   RR$_{c}$    &0.70 (C99) &		         &  14.913& \\
V114  &  0.604015   (O99) &  47629.573   (C99) &  13.794  &  0.003 &   RR$_{ab}$   &1.08 (C99) &		         &  15.120 &    \\
V115  &  0.6034     (C99) &  47629.772   (C99) &  13.639  &  0.003 &   RR$_{ab}$   &0.77 (C99) &		         &  14.954 &no PLK/PLJ-crowded\\
V116  &  0.347421   (O99) &  47629.518   (C99) &  14.071  &  0.003 &   RR$_{c}$    &0.59 (C99) & +0.40	    &  14.787    &     \\
V117  &  0.335578   (O99) &  47629.520   (C99) &  14.077  &  0.003 &   RR$_{c}$    &0.53 (C99) & -0.48	    &  14.895     &          \\
V118  &     0.5805        (C99) &     47629.832      (C99) &    13.454     &     0.004    &  RR$_{ab}$   &  1.57    (C99)&-0.10&   14.693    &  no PLK/PLJ\\
V119  &  0.5629     (C99) &  47629.297   (C99) &  13.783  &  0.003 &   RR$_{ab}$   &1.25 (C99) & +0.20	    &  14.935              &  \\
 V120  &     0.2797        (C99) &     47629.698      (C99) &     14.474     &     0.004    &  RR$_{c}$    &   0.71    (C99)&+0.10&  15.018     &  no PLK/PLJ-crowded\\    
V121  &  0.623304   (B96) &  47629.303   (C99) &  13.733  &  0.003 &   RR$_{ab}$   &1.69 (C99) & +0.05	    &  14.846     &     \\
V122  &  0.58048    (Cl)&        51946.869*    & 13.47**  &  &   RR$_{ab}$  &  & &14.611       & no PLK/PLJ\\
V123  &  0.6025     (C99) &  47629.303     &  13.715  &  0.003 &   RR$_{ab}$   &0.65 (C99) & -0.10	    &  14.989                & \\
 V125  &     0.3065        (C99) &     47629.592      (C99) &        13.868     &     0.005    &  RR$_{c}$    &    0.65    (C99)& &  14.933           & no PLK/PLJ-crowded \\
 V127  &     0.544965      (O99)&     51946.869*     &    13.80**     &    &  RR$_{ab}$   &  & &   15.059          &    no PLK/PLJ \\
V128  &  0.305704   (O99) &  47629.785   (C99) &  14.240  &  0.003 &   RR$_{c}$    &0.68 (C99) &		         &  14.904   &    \\
V129  &  0.6011     (C99) &  47629.631   (C99) &  13.783  &  0.003 &   RR$_{ab}$   &0.74 (C99) & 	         &  15.112              & \\
V130  &  0.327396   (R96) &  47581.469   (C99) &  14.100  &  0.003 &   RR$_{c}$    &0.50 (R96) & +0.46	    &  14.887                 &\\
V131  &  0.281521   (R96) &  47629.763   (C99) &  14.286  &  0.003 &   RR$_{c}$    &0.68 (C99) & +0.30	    &  15.157                &  \\
V132  &  0.2835     (C99) &  47629.648   (C99) &  14.255  &  0.004 &   RR$_{c}$    &0.49 (C99) & +0.20	    &  14.974              &\\
 V133  &     0.294905      (Cl99)&   52331.804*     &      14.14**     &     &  RR$_{c}$   &  & &   14.869          &   no PLK/PLJ \\
 V137  &     0.591         (C99) &     47629.520      (C99) &      13.825     &     0.003    &   RR$_{ab}$  &  0.60    (C99)& +0.20&  15.020      & \\
V139  &  0.300      (C99) &  47629.587   (C99) &  14.186  &  0.003 &   RR$_{c}$    &0.42 (C99) &		         &  14.946    &   \\
V142  &  0.4577     (C99) &  47629.830    (C99) &  14.078  &  0.003 &   RR$_{ab}$   &1.5  (C99) &		         &  15.194 &no PLJ\\
V155  &    0.33    (DS)&     52331.827*      &     14.101   &            &  RR$_{c}$    &    & & 14.882          &    no PLK/PLJ\\
V156  &    $>$0.47    (DS)&     51946.869*     &        12.73**    &      &  RR$_{ab}$   & & &    14.984          &    no PLK/PLJ-blended? \\
V158  & 0.45  (DS) &      51947.004*      &  13.58** &   &   RR$_{ab}$    & & & 14.235       & no PLK/PLJ\\
 V161  &    0.331570     (O99)&    51946.736*         &     14.035    &        &   RR$_{c}$  &   & &  15.051          &   no PLK/PLJ    \\
 V162   &    0.557468     (O99)&      51946.925*      &       13.86**     &         &   RR$_{ab}$  &  &  &     15.060          & no PLK   \\
 V163 &    0.600853     (O99)&         52331.773* &   13.86**     &      &   RR$_{ab}$  &   &  &     14.912          &     no PLK \\
\hline 
\multicolumn{7}{l}{\footnotesize{$*$ Epoch of our first measurement.}}\\
\multicolumn{7}{l}{\footnotesize{$**$ Intensity-weighted mean magnitude computed by DAOMASTER.}}\\
\multicolumn{7}{l}{\footnotesize{References:}}\\
\multicolumn{7}{l}{\footnotesize{S91:~\cite{stormetal_1991PASP}.}}\\
\multicolumn{7}{l}{\footnotesize{B96:~\cite{brocatoetal_1996AJ}.}}\\
\multicolumn{7}{l}{\footnotesize{R96:~\cite{reid_1996MNRAS}.}}\\
\multicolumn{7}{l}{\footnotesize{DS:~\cite{drisse&shara_1998AJ}.}}\\
\multicolumn{7}{l}{\footnotesize{C99:~\cite{caputoetal_1999MNRAS}.}}\\
\multicolumn{7}{l}{\footnotesize{O99:~\cite{olechetal_1999MNRAS}.}}\\
\multicolumn{7}{l}{\footnotesize{K00:~\cite{kaluznyetal_2000A&AS}.}}\\
\multicolumn{7}{l}{\footnotesize{S10:~\cite{szeidl_2010arXiv1010}.}}\\
\end{tabular}
\end{minipage}
\end{table*}

\subsection{Notes on individual variables}\label{sec_noterrl}
\textbf{V25}: for this variable we did not find any amplitude information in the literature. However since the pulsation cycle was evenly covered, we fitted the observed data with a spline curve, getting a light curve (red curve in Fig.~\ref{fig_curve}) with a shape very similar to the template available for similar periods. \\
\textbf{V84}: actually this variable is a type II Cepheid, whose $K$-band light curve was presented by \citet{matsunaga2006}, and it is included in our photometric catalog. Unfortunately, most of our measurements of this star are above the non-linearity level, and we were not able to produce a reliable light curve.\\
\textbf{V86}: this variable is heavily contaminated by close neighbors.\\
\textbf{V96}: this variable is heavily contaminated by close neighbors.\\
\textbf{V104}: this variable has been claimed to be a RR$_{ab}$ variable by~\cite{reid_1996MNRAS}, but~\cite{drisse&shara_1998AJ} reported it as an eclipsing binary. We found that this variable does not fit the PLK relation (see Fig.~\ref{fig_plkb01}), since its magnitude is $\sim 0.2$ mag brighter than the expected magnitude for its period, while the observed dispersion of the PLK relation is $\sim 0.15$ mag at the $ 3 \sigma$ level. This  gives support to the~\cite{drisse&shara_1998AJ} interpretation, as also stated in~\cite{caputoetal_1999MNRAS}. However,~\cite{olechetal_1999MNRAS} reported this star as a multiperiodic variable, with a first overtone radial pulsation and a second period connected with non-radial pulsation.\\
\textbf{V106}: this variable is heavily contaminated by close neighbors. \\
\textbf{V107}: this variable is heavily contaminated by close neighbors.\\
\textbf{V108} this variable appears to be overluminous in the PLK plane (see Fig.~\ref{fig_plkb01}), and both overluminous and redder than expected from its pulsation period in the CMD (magenta point in Fig.~\ref{fig_cmd}). We therefore suspect that this variable is blended with a red companion, and we excluded it from our analysis.\\
\textbf{V111}: this variable is heavily contaminated by close neighbors.\\
\textbf{V113}: according to~\cite{caputoetal_1999MNRAS} this variable shows a ($B-V$) color significantly bluer than the blue limit of the instability strip, but with a luminosity which agrees with the average luminosity of the other variables. Excluding subtle effects of blending or crowding, the blue color remains unexplained. However, we found that this variable fairly matches the PLK relation and we used it in our analysis.\\
\textbf{V115}: this variable is heavily contaminated by close neighbors.\\
\textbf{V118}: this variable appears to be over-luminous in the PLK plane (green point in Fig.~\ref{fig_plkb01}), and it has not been used for our PLK analysis. Since it does not seem to be blended, its over-luminosity in the $K$-band is unexplained. We therefore dropped this star from the present study.\\
\textbf{V120}: this variable is heavily contaminated by close neighbors.\\
\textbf{V125}: this variable is heavily contaminated by close neighbors.\\
\textbf{V156}: this variable appears both overluminous and red in the CMD. We therefore suspected that this variable is blended with a red companion, and we excluded it from our analysis. In Fig. \ref{fig_curve} we plot this variable after shifting the $K$ magnitude by 1 mag, to preserve the readability of the figure.

\begin{figure*}
\includegraphics[width=18cm,height=20cm]{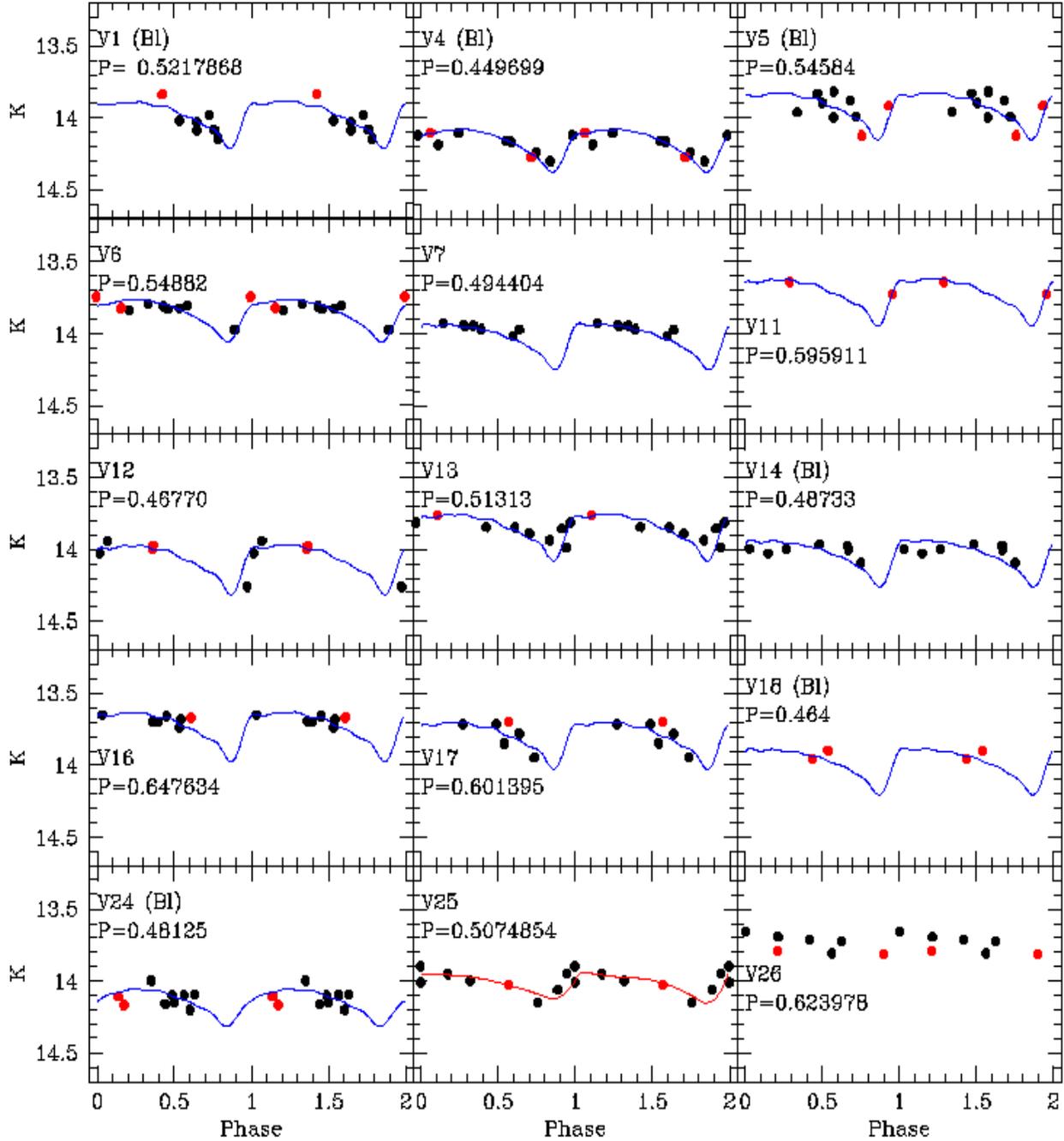}
\caption{$K$-band light curves for all RRLs in our fields. Black and red points mark SOFI and NICS data, respectively.
The blue solid lines show, when available, the fit with the template light curve, while the red line the fit 
with a spline.}
\label{fig_curve}
\end{figure*}
\begin{figure*}
\includegraphics[width=18cm,height=20cm]{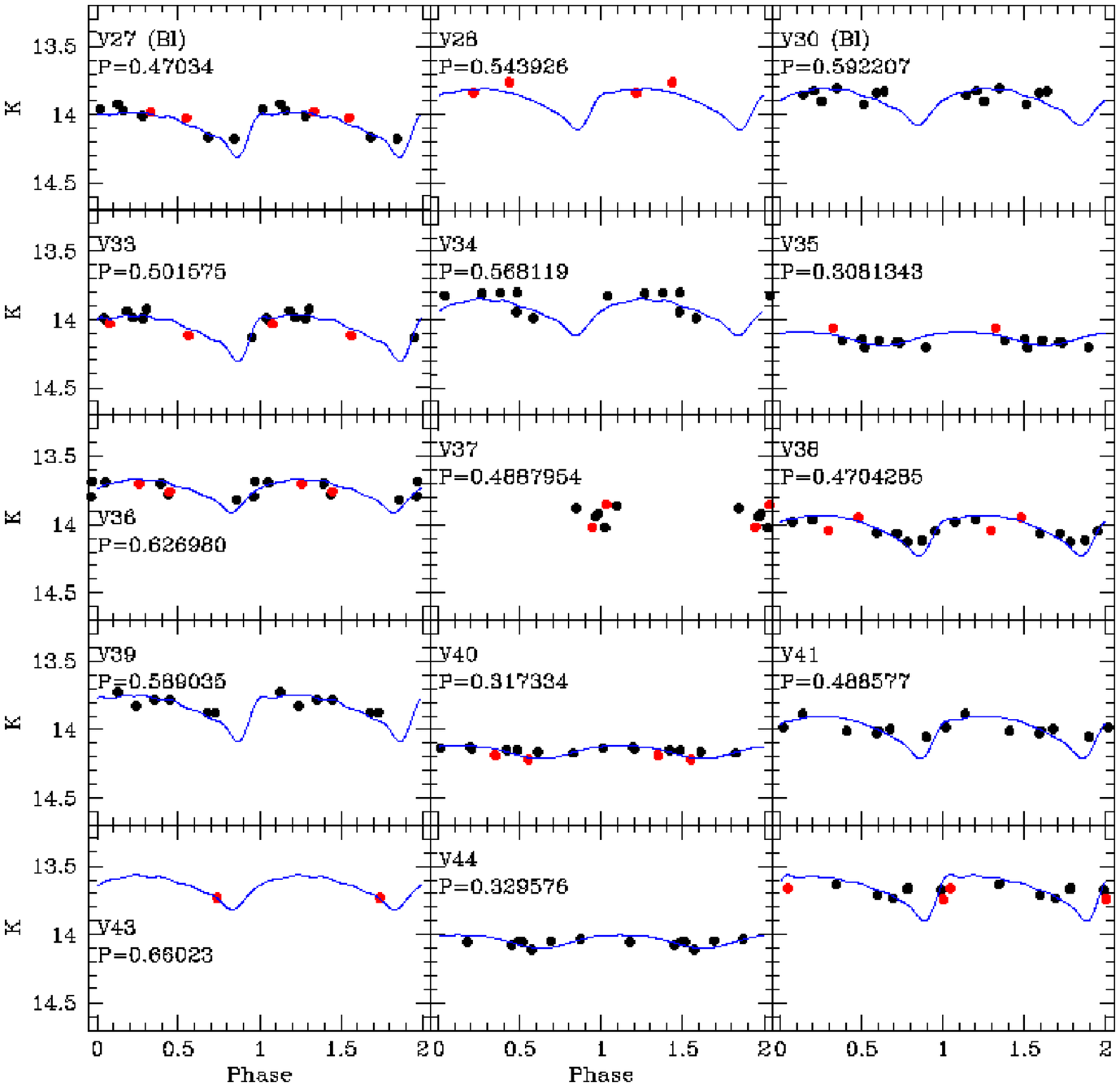}
\contcaption{}
\end{figure*}
\begin{figure*}
\includegraphics[width=18cm,height=20cm]{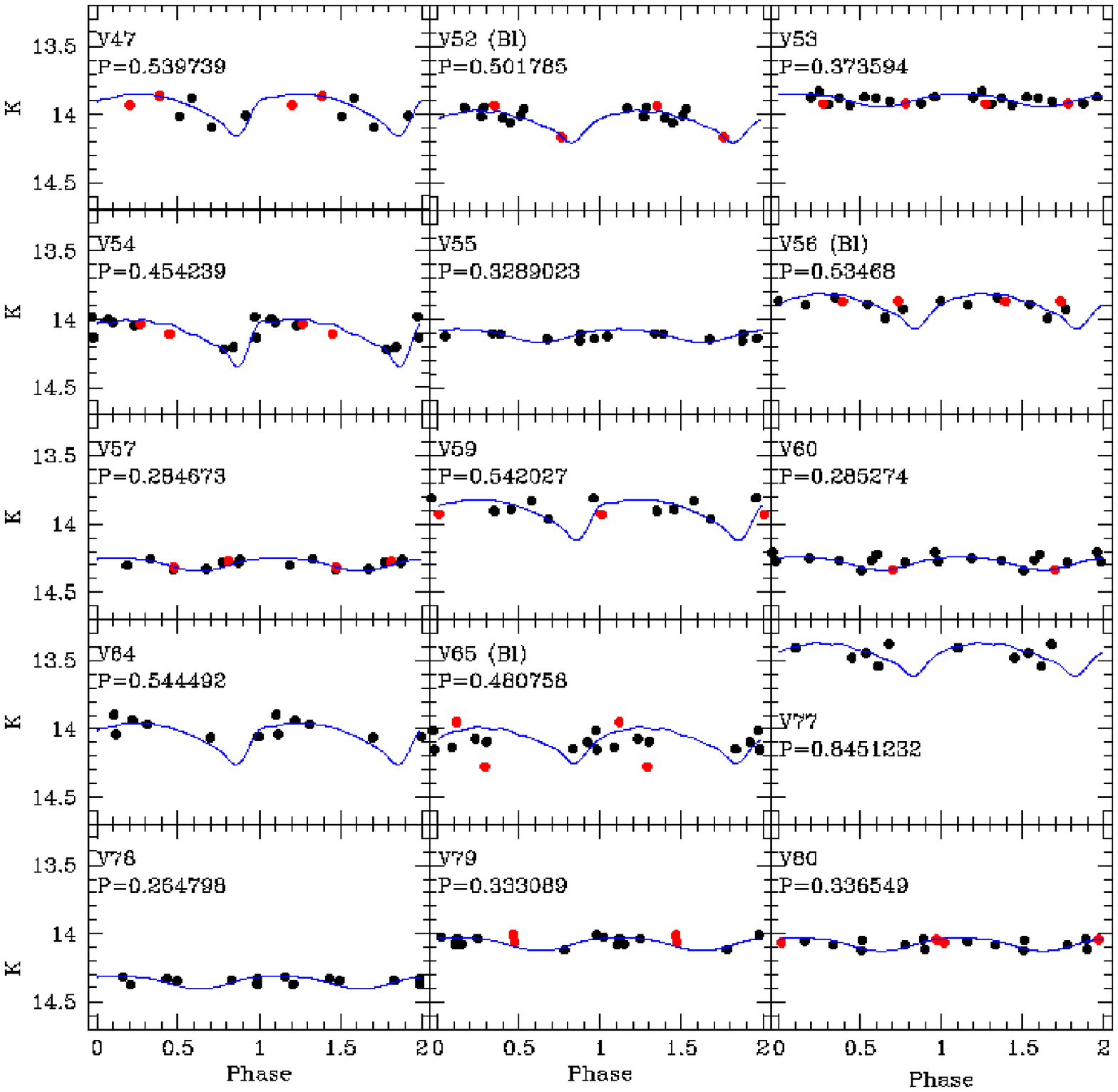}
\contcaption{}
\end{figure*}
\begin{figure*}
\includegraphics[width=18cm,height=20cm]{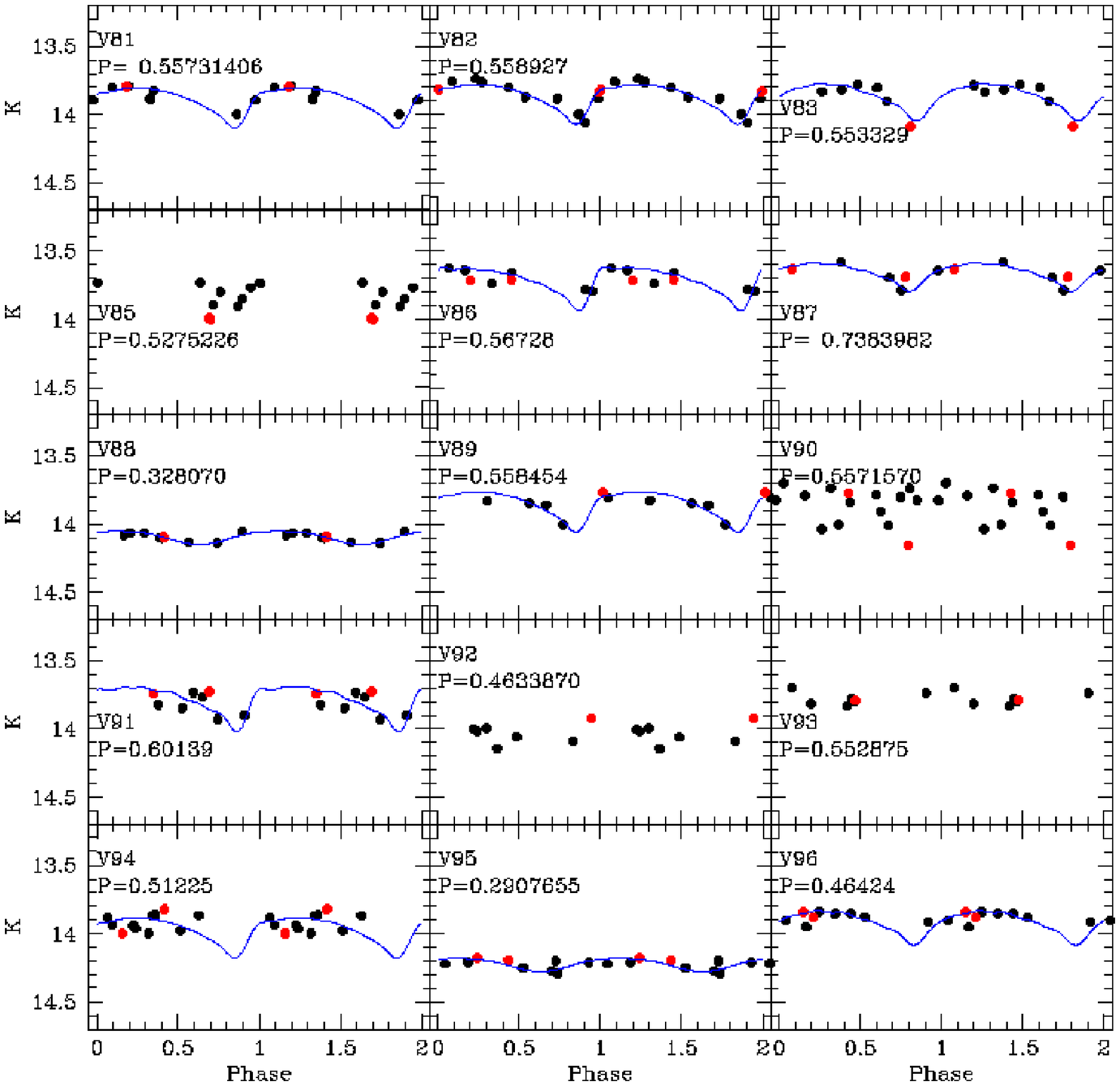}
\contcaption{}
\end{figure*}
\begin{figure*}
\includegraphics[width=18cm,height=20cm]{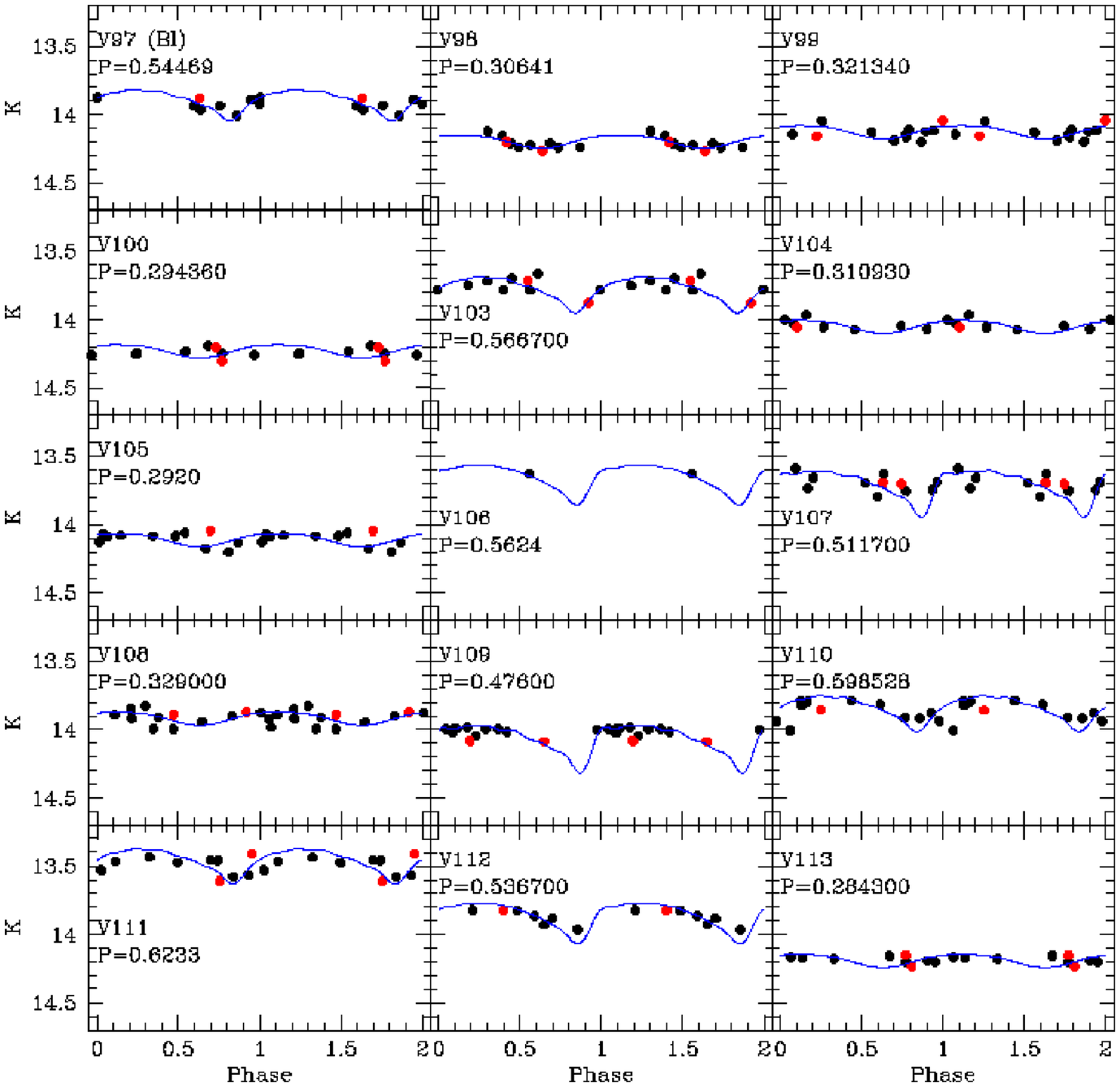}
\contcaption{}
\end{figure*}
\begin{figure*}
\includegraphics[width=18cm,height=20cm]{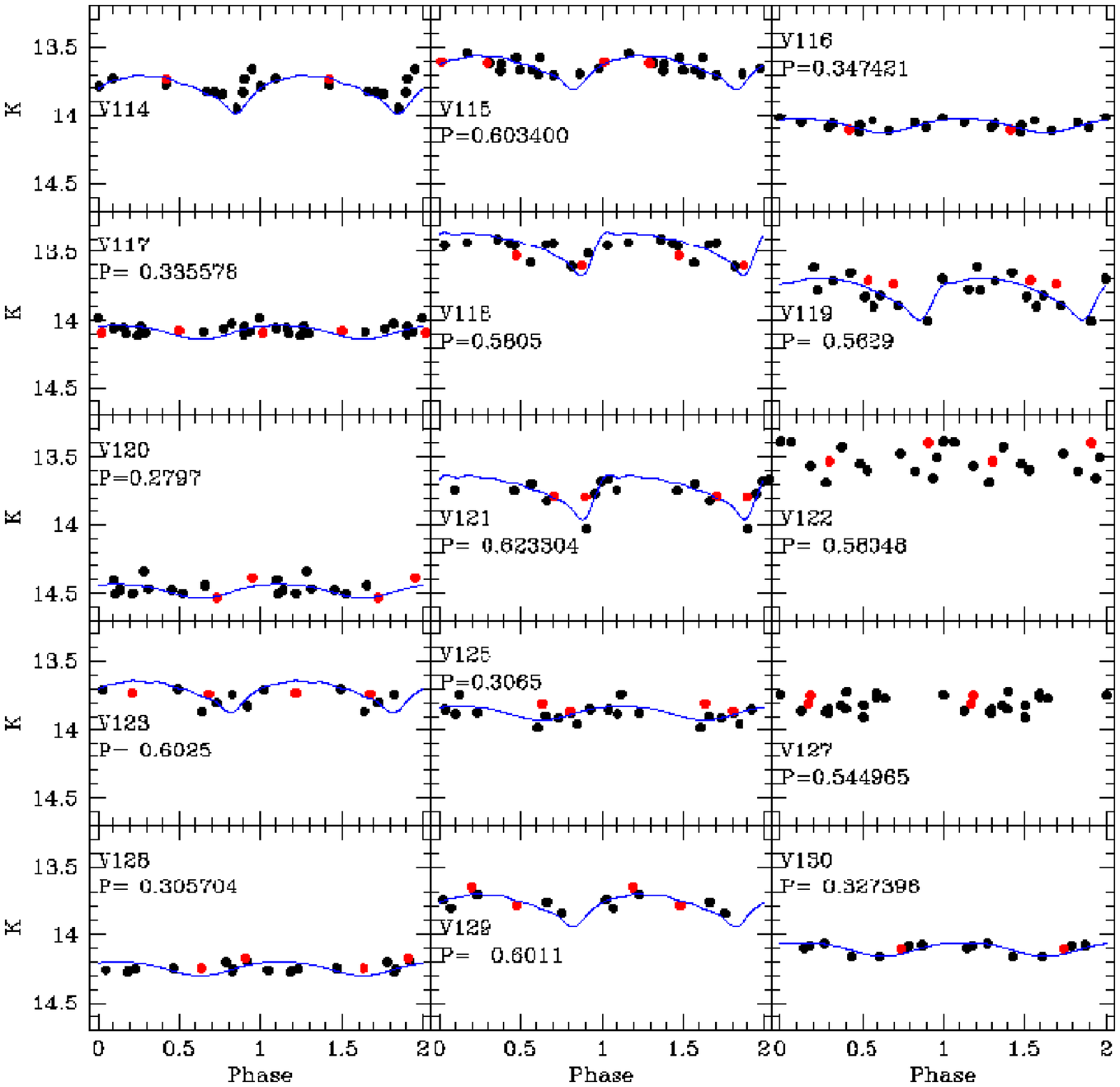}
\contcaption{}
\end{figure*}
\begin{figure*}
\includegraphics[width=18cm,height=20cm]{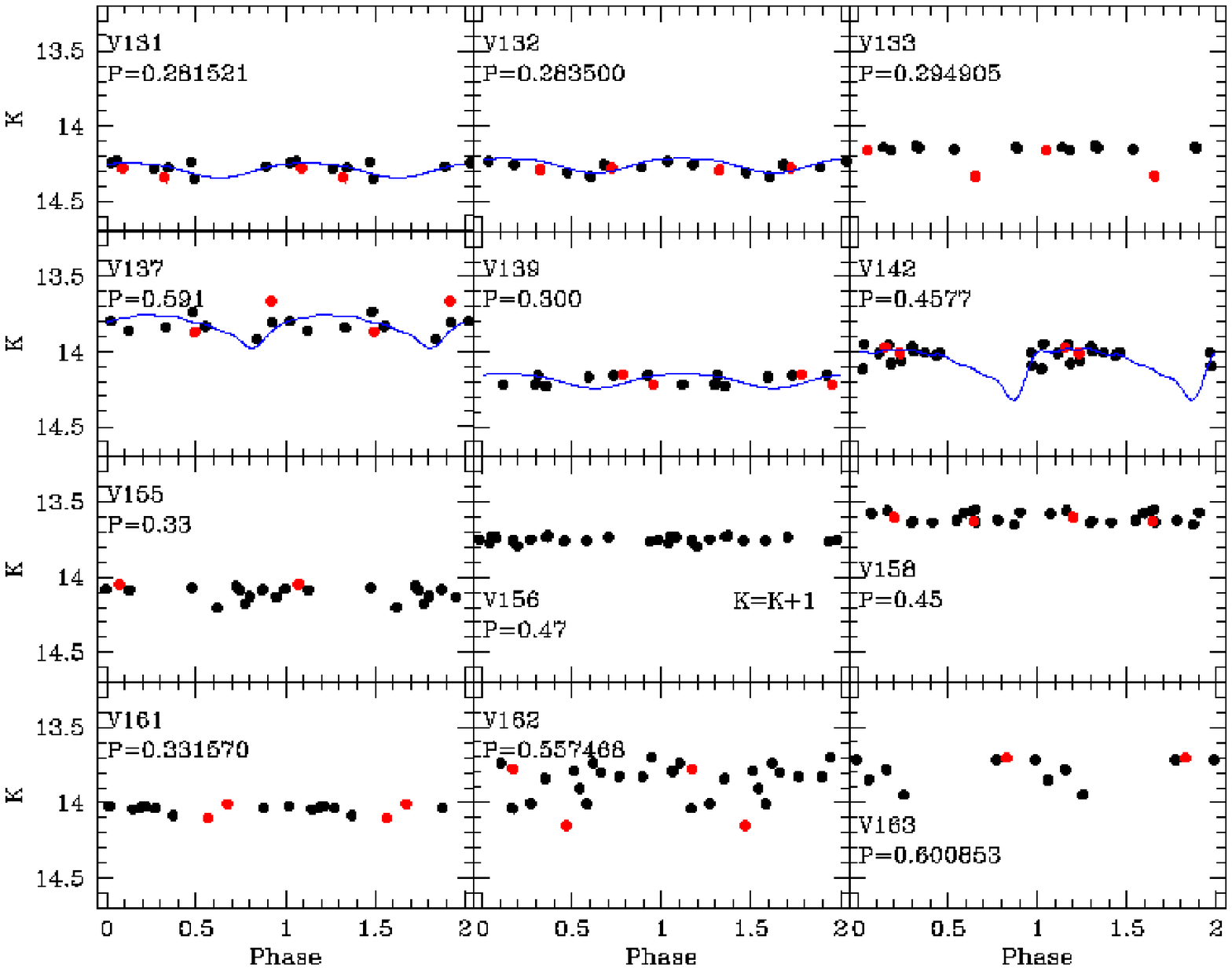}
\contcaption{The $K$-band magnitudes of V156 have been artificially shifted by 1 mag to make more clear the figure.}
\end{figure*}

\section{The PLK and the PLJ relations}\label{sec_pl}
In the following we compare our data with the available PLK calibrations, 
which can be divided into three broad groups: pulsational (\citealt{bonoetal_2001mnras};~\citealt{bonoetal_2003mnras}), 
synthetic (\citealt{cassisietal_2004AA}; \citealt{catelanetal_2004ApJS}) and empirical (\citealt{sollimaetal_2006mnras}) 
PLK relations. We also compare our data with the PLJ calibration provided by \citealt{catelanetal_2004ApJS}.

\begin{figure*}
\includegraphics[width=9cm,height=9cm]{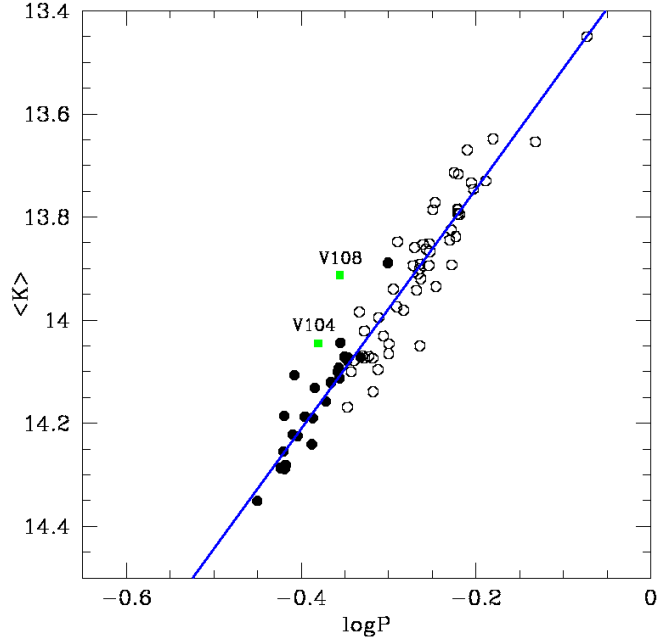}
\caption{Empirical $\log P-<K>$ relation of the M5 RRLs. Filled circles show the fundamentalized
RR$_{c}$ stars. Open circles are RR$_{ab}$ stars, and the solid blue line represents the empirical fit to the data.}\label{fig_plkb01}
\end{figure*}

\begin{figure*}
\includegraphics[width=9cm,height=9cm]{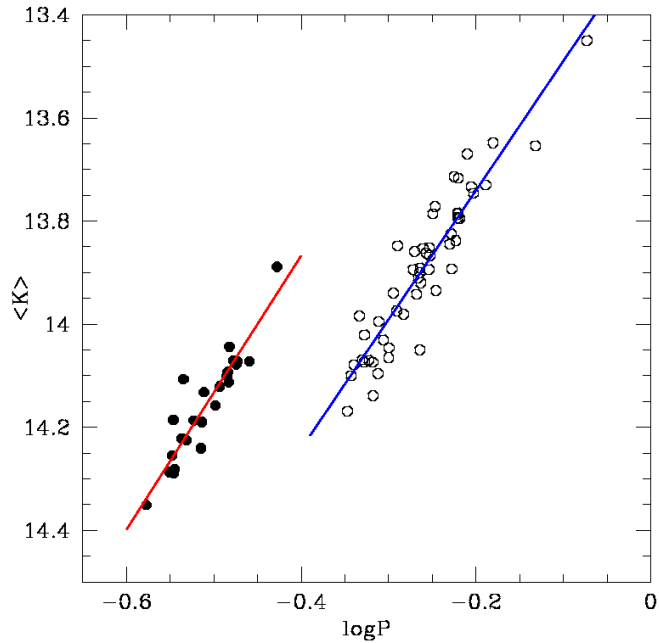}
\caption{Observed $\log P-<K>$ relation of the M5 RRLs. Filled circles show the RR$_{c}$ stars, while open circles are RR$_{ab}$ stars. Solid blue and red lines represent the empirical fit to the data.}\label{fig_plkb03}
\end{figure*}

Fig.~\ref{fig_plkb01} shows the observed mean $K$-band magnitudes of the detected variables 
as a function of the period (observed PLK relation), the empirical fit (rms$=0.05$) to the data (blue line) is
\begin{equation}\label{eq_bono01}
 	\langle K\rangle=-2.33(\pm 0.08)\log P + 13.28(\pm 0.02),
\end{equation}
obtained \textit{fundamentalizing} the
first overtone pulsators (filled circles) by applying the relation $\log P_F\, =\, 0.127 + \log P_{FO}$~\citep{dicriscienzoetal_2004apj}, to use the same PLK relation for 
fundamental (open circles) and first overtone pulsators. We remark that this offset is consistent with that implied by the difference in $\log P$ at costant $\langle K\rangle$ in Fig. \ref{fig_plkb01}. The slope in Eq.~\ref{eq_bono01} is consistent within $1 \sigma$ with the empirical values found for this cluster ($-2.42 \pm 0.23$  \citealt{longmoreetal_1990mnras}; $-2.27$ \citealt{sollimaetal_2006mnras}, unfortunately without uncertainty). Observed mean $K$-band magnitudes of R$_{ab}$ and RR$_{c}$ are shown separately as a function of the period in Fig.~\ref{fig_plkb03}. Symbols are the same of Fig.~\ref{fig_plkb01}. Solid blue and red lines show the empirical fits to the RR$_{ab}$ and RR$_{c}$ variables, respectively, in particular for the RR$_{ab}$ variables:
\begin{equation}\label{eq_bono03f}
 \langle K\rangle =-2.50(\pm 0.14)\log P + 13.24(\pm 0.04)
\end{equation} 
and for RR$_{c}$:
\begin{equation}\label{eq_bono03fo}
 \langle K\rangle =-2.66(\pm 0.23)\log P + 12.8(\pm 0.1).
\end{equation}

By adopting the theoretical calibration of~\citet{bonoetal_2001mnras} (their eq. $[2]$) and a 
metallicity of $[Fe/H]=-1.26$~\citep{kraft&ivans_2003PASP}, we can determine the individual distance 
moduli of the RRLs. The observations were transformed into the \citet{Bessel88} homogenized Johnson-Cousins-Glass 
photometric system according to the relation $K_{BB} = K_{2M} +0.044 \, \rm mag$ provided by \citet{carpenter2001}. 
The weighted average of these estimates gives an apparent distance modulus $\rm DM_K=(14.42 \pm 0.06) \, \rm mag$, where the adopted uncertainty is the standard deviation (rms). The adopted reddening toward M5 is $E(B-V)=0.03\, \rm mag$, as taken from the \cite{harris_2010aj} catalog. We combine this value with the reddening law from~\cite{cardellietal_1989ApJ} of $A_{K}=0.114 \times 3.1 \times E(B-V)$ 
which gives an absorption in the $K$-band of $A_{K}=0.011$ mag. Correcting the distance modulus for 
the reddening, we find that the true value is $\rm DM_0=(14.41 \pm 0.06) \, \rm mag$. These results indicate that 
the current distance estimate to M5 is minimally affected by reddening uncertainties and by uncertainties in
the extinction law (\citealt{mccall_2004aj,bonoetal_2010ApJ}).

In order to improve the theoretical calibration of the PLZ$_{K}$ relation,~\cite{bonoetal_2003mnras} devised 
a new pulsation approach that relies on mean $K$-band magnitudes and ($V-K$) colors. In particular, 
they derived new period-luminosity-color-metallicity relations for RR$_{ab}$ and 
RR$_{c}$ variables (see their relations $[7]$ and $[8]$) that 
include the luminosity term. According to these relations, one finds that accurate $V$, $K$ photometric measurements of both RR$_{ab}$ and RR$_{c}$ variables provide excellent proxies for the effective temperature, and can in turn be adopted to estimate the luminosity level.
We therefore adopted accurate $V$-band magnitudes extracted by optical light curves of the PBS unpublished archive (shown in the last column of Table~\ref{tab_rrl}) and we compared our empirical slopes in Eqs.~\ref{eq_bono03f} and~\ref{eq_bono03fo} with the theoretical predictions, being $-2.102$ and $-2.265$ for RR$_{ab}$ and RR$_{c}$ stars, respectively.
Once we transform the observations into the \citet{Bessel88} system and after correcting for the reddening, 
we find a value of $\rm DM_0=(14.46 \pm 0.09 \, rms) \, \rm mag$ using the RR$_{ab}$ stars, while using the RR$_{c}$ stars we find $\rm DM_0=(14.46 \pm 0.05 \, rms) \, \rm mag$. Note that the dispersion of the distance estimates based on RR$_{c}$ stars is almost a 
factor of two smaller than the distance based on RR$_{ab}$ stars, because the width in temperature of 
the region of the instability strip in which the FOs are pulsationally stable is typically a factor of 
two narrower than for the Fs. This means that the intrinsic dispersion in the infrared luminosity of FOs is 
systematically smaller than for Fs.  

The evolutionary calibrations by ~\cite{cassisietal_2004AA} and~\cite{catelanetal_2004ApJS} are based on 
the Horizontal Branch morphology and on the RRL pulsational properties. These authors provide 
period-luminosity relations in the Johnson-Cousins-Glass photometric system, with slopes and zero-points 
that vary with the HB morphology and with the metallicity.
In particular, ~\cite{cassisietal_2004AA} gives a grid of slopes and zero-points as a function of the 
metallicity (expressed as the mass fraction $Z$) and the HB type (HBT, according to the Lee, 
Demarque \& Zinn parameter, \citealt{Lee94}). We chose, among the values available in their published grid, $Z=0.0006$ and HBT $= 0.28$ since these parameters are those closest to the  M5 literature values
(HBT $=0.31$, taken from the \citealt{harris_2010aj} catalog), obtaining the calibration $\rm M_{K} = -2.34 (log P_{F} + 0.30 )-0.394$. 
Once we apply the correction for the difference in the photometric system and for the reddening, and having fundamentalized
RR$_c$ variables, we find a true distance modulus to M5 of $\rm DM_0=(14.41 \pm 0.05 \,rms) \, \rm mag$.

A very similar approach was devised by \citet{catelanetal_2004ApJS} who provided \textit{average} relations, 
to be used when the HB type is not known {\it a priori\/}. However, we adopted their calibration with $Z=0.0005$ and 
HBT $= 0.414$, obtaining the calibration $\rm M_{K}\, =\, -2.355 (\log P_{F})-1.172$. 
Again, once we apply the correction for the difference in the photometric system and for the reddening, and having fundamentalized 
the RR$_c$ variables, we found a true distance modulus to M5 of $\rm DM_0=(14.48 \pm 0.05 \, rms) \, \rm mag$.

Finally, we also adopted the empirical calibration by~\cite{sollimaetal_2006mnras} 
($\rm M_{K} \,=\, -2.38 \log P_{F} + 0.09\, [Fe/H] -1.04$) and we found a true distance modulus of $\rm DM_0=(14.41 \pm 0.05 \, rms) \, \rm mag$. It is interesting to note that~\cite{sollimaetal_2006mnras}, using a collection of data for $86$ RRLs, found a true distance to M5 ($14.35 \pm 0.15 \, \rm mag$) that agrees well with the current estimates.

\begin{figure*}
\includegraphics[width=9cm,height=9cm]{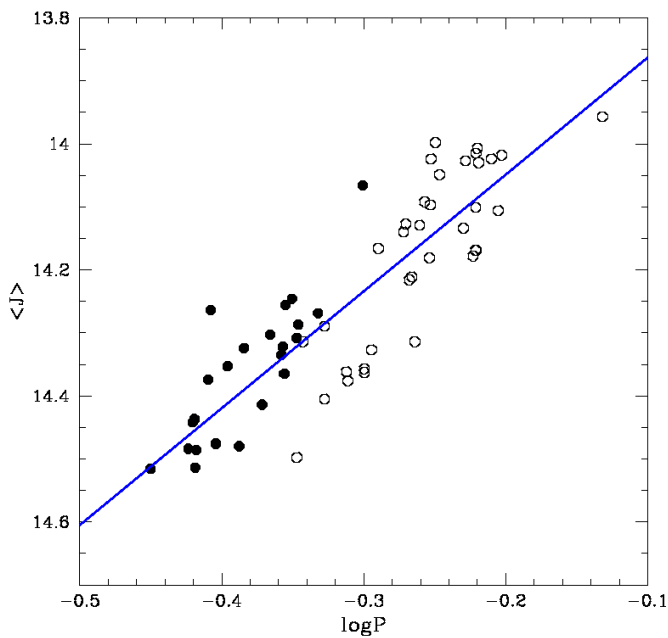}
\caption{Observed $\log P-<J>$ relation of the M5 RRLs. Filled circles show fundamentalized RR$_{c}$ stars, while open circles are RR$_{ab}$ stars. Solid blue line represents the empirical fit to the data.}\label{fig_plj}
\end{figure*}

We end this section with a discussion of the observed PLJ relation. Fig.~\ref{fig_plj} shows observed 
mean $J$ magnitudes of the considered RRLs vs. $\log P$. The $RR_c$ variables have been fundamentalized. 
Since in the literature there are no suitable templates to adopt, and since the uneven phase coverage of the 
observations did not allow us to interpolate the observed data with spline curves, we simply used 
intensity-weighted mean magnitudes, as computed with DAOMASTER. To reduce subtle effects for uneven light curve sampling, we used for the PLJ only stars with a good phase coverage, therefore ending up with $35$ RR$_{ab}$ and $23$ RR$_c$ stars. The observed slope is $-1.85 \pm 0.14$, which is in very good agreement with the available theoretical calibration ($\rm M_{J} \,=\, -1.773 \log P + 0.190 \log Z -0.141$,~\citealt{catelanetal_2004ApJS}). In passing, we note that F and FO pulsators seem to not follow a common relation even after fundamentalization, with the F variables alone apparently following a steeper distribution. On the other hand, we point out that the~\citealt{catelanetal_2004ApJS} calibration used as reference is obtained by using both $RR_{ab}$ and $RR_c$ stars.\\
We transformed the observations in the \citet{Bessel88} homogenized Johnson-Cousins-Glass $J$ system with the relation $J_{BB} = J_{2M} -0.029 (J-K)_{2M} +0.053$, derived following the \citet{carpenter2001} equations. After correcting for the extinction with the~\cite{cardellietal_1989ApJ} law, we get a true PLJ-based distance to M5 of $(14.50 \pm 0.08) \, \rm mag$, in agreement with the PLK calibrations. We explicitly note that the distances based on the two PLK and PLJ calibrations provided by \citet{catelanetal_2004ApJS} are slightly longer than the others, but with an excellent internal agreement.

All predicted and empirical slopes and the distance moduli based on the above relations are listed in Table~\ref{tab_dist_mod}.

\begin{table*}
\caption{NIR PL slope and true distance moduli for different adopted calibrations. F and FO labels refer to fundamental and first overtone variables, respectively. The adopted error is the rms of the distribution of the estimates on individual RRLs.}\label{tab_dist_mod}
\begin{tabular}{|l|c|c|c|c|c|}
\hline 
Calibration & Predicted slope & Sample &Our Empirical $J$ slope & Our Empirical $K$ slope &  \rm DM$_0$ (mag)\\
      \hline 
\cite{bonoetal_2001mnras} & -2.031 & F $+$ FO & -- & -2.33 $\pm$ 0.08 & 14.41 $\pm$ 0.06\\
\cite{bonoetal_2003mnras} & -2.102 & F & --- & -2.50 $\pm$ 0.14 & 14.46 $\pm$ 0.09\\
\cite{bonoetal_2003mnras}& -2.265 & FO  & --- & -2.66 $\pm$ 0.23 & 14.46 $\pm$ 0.05\\
\cite{cassisietal_2004AA}&-2.34& F$+$FO & --- & -2.33 $\pm$ 0.08 & 14.41 $\pm$ 0.05\\
\cite{catelanetal_2004ApJS}&-2.355&F$+$FO& ---  & -2.33 $\pm$ 0.08 & 14.48 $\pm$ 0.05\\
\cite{catelanetal_2004ApJS}&-1.773& F$+$FO & -1.85 $\pm$ 0.14  & --- & 14.50 $\pm$ 0.08\\
\cite{sollimaetal_2006mnras}&-2.38 & F$+$FO& --- & -2.33 $\pm$ 0.08 & 14.41 $\pm$ 0.05\\
\hline 
\end{tabular}
\end{table*}

\begin{table*}
\caption{Summary of some of the M5 distance estimates available in the literature.}\label{tab_dist_mod_all}
\begin{tabular}{|l|l|l|l|l|}
\hline 
Study & Method & [Fe/H]&  Reddening & \rm DM$_0$ (mag)\\
      \hline 
\cite{sollimaetal_2006mnras}    & PLK                     & $-1.10$ & $E(B-V)=0.035$ & $14.35 \pm 0.15$\\
\cite{reid_1998AJ}              & Main sequence fitting   & $-1.10$ & $E(B-V)=0.02$ & $14.52 \pm 0.15$\\
\cite{carrettaetal_2000ApJ}     & Main sequence fitting   & $-1.10$ & $E(B-V)=0.035$ & $14.48 \pm 0.05$\\
\cite{layden2005}               & Main sequence fitting   & $-1.11$  &$E(V-I)=0.046$ & $14.45 \pm 0.11$\\
\cite{stormetal_1994AA}        & Baade-Wesselink         & $-1.40$  &  $E(V-I)=0.02$ & $14.37 \pm 0.18$\\
\cite{dicriscienzoetal_2004apj} & RRL pulsation properties & $-1.26$ &  $E(B-V)=0.03$ & $14.32 \pm 0.04$\\
\cite{rees_1996}                & Proper motions          &\ldots &\ldots & $14.44 \pm 0.41$\\
\cite{layden2005}               & White dwarf fitting     & $-1.11$  &$E(V-I)=0.046$ & $14.67 \pm 0.18$\\
\hline 
\end{tabular}
\end{table*}

\section{Discussion}\label{sec_con}

On the basis of our data, we found that the weighted mean of the different theoretical and empirical calibrations of the PLK relation gives a true distance modulus of $14.44 \pm 0.02$ mag.

The distance to M5 based on the PLK relation also agrees quite well with similar estimates available in the literature, but based on different distance indicators. The distance estimates listed in Table~\ref{tab_dist_mod_all} indicate quite clearly that the current distance agrees within $1 \sigma$ with distance based not only on the PLK relation~\citep[$14.35\pm0.15$ mag][]{sollimaetal_2006mnras}, 
but also on main-sequence fitting method \citep{reid_1998AJ,carrettaetal_2000ApJ,layden2005} and semi-geometric approach~\citep[$14.37\pm0.18$ mag][]{stormetal_1994AA}. On the other hand, our distance is systematically longer than the distance to M5 provided by~\citet{dicriscienzoetal_2004apj} using 
synthetic horizontal branch models and pulsation properties of RRLs and the difference 
is larger than 1$\sigma$.

In this context it is noteworthy that the distance based on the PLK relation 
agrees with the distance based on the proper motions $(m-M)_0$=$14.44 \pm 0.41$ mag by~\citet{rees_1996}. 
Even though the intrinsic error of the quoted distance estimates differ by more than one order 
of magnitude, the two distances taken at face value agree quite well. If this agreement is not fortuitous, due to the large error bar in the ~\citet{rees_1996} estimate, this would be first time that kinematic distances agree with other distance indicators. Typically, the kinematic distances to GCs 
are 0.2-0.3 magnitudes smaller than distance moduli based either on the PLK relation of RRLs, 
or on main-sequence fitting, or on the tip of the Red Giant Branch (TRGB,~Bono et al.~\citeyear{bonoetal_2008ApJ}).  
The reasons for such discrepancies are not clear yet, but the current agreement appears very 
promising to further constrain the occurrence of possible systematic errors.

On the other hand, the distance based on the fitting of the White dwarf cooling sequence 
$(m-M)_0$=$14.67 \pm 0.18$ mag provided by \cite{layden2005} is approximately 0.2 mag 
larger than distances based on other methods. The difference with the distance based on 
the PLK relation and on main-sequence fitting is on average slightly larger than 1$\sigma$.    
This is also an interesting occurrence, since distance moduli to GCs based on the fitting 
of the white-dwarf cooling sequence are typically 0.1-0.3 mag smaller than distances 
to GCs based either on the TRGB, or on the PLK relation or on the main sequence fitting~\citep{bonoetal_2008ApJ,moehler2008arXiv}.  

We have also investigated the PLJ relation of RRL stars in M5, and from a sample of 35
RRab and 23 RRc (fundamentalized) variables we found an empirical slope of $-1.85 \pm 0.14$. 
The uncertainty in the slope of the PLJ relation is almost a factor of two 
larger than the uncertainty in the slope of the PLK relation (0.14 vs 0.08) because the
luminosity amplitude in the $J$-band is larger in the $K$-band, and also because we 
still lack accurate $J$-band light curve templates. Therefore, the mean $J$-band magnitudes 
were estimated as a time average and only for RRLs with good phase coverage. Homogeneous 
sets of time series data both in $J$ and $K$ are required to constrain 
on an empirical basis the intrinsic spread of PLK and PLJ relations. Finally, we mention 
that by adopting the calibration of the PLJ relation provided by ~\citet{catelanetal_2004ApJS}, 
we found a true distance modulus to M5 of $14.50 \pm 0.08$ mag. The distances based on the 
PLJ and on the PLK relations agree quite well, but the intrinsic error of the former is
a factor of two larger. 

The distances to M5 based either on the PLK or on the PLJ relation agree quite well  
with distances based on independent robust standard candles. This finding further supports 
the key role that accurate NIR photometry of cluster variables (RRL, type II Cepheids) can play 
in the improvement of GC distance scale \citep{matsunaga2010}. Moreover and even more importantly, 
the comparison with M5 distances based on different distance indicators strongly supports the evidence 
that M5 might be a fundamental laboratory to constrain on a quantitative basis the thorny systematic 
uncertainties that might affect the most popular primary distance indicators.    
It goes without saying that this sanity check would benefit by further improvements in the 
precision of the kinematic distances and/or in the geometrical distances based on possible 
eclipsing binary systems in M5. 

In a future paper we plan to compare optical and NIR photometry of M5 with evolutionary 
predictions, in particular for advanced evolutionary phases.  

\section*{Acknowledgments}
We sincerely thank an anonymous referee for her/his helpful comments, which improved the readibility of the paper.\\
GC is supported by the Italian Ministry of Education, University and Research (MIUR) grant PRIN-MIUR 2007:\textit{Multiple stellar populations in globular clusters: census, characterization and origin} P.I.: G. Piotto.\\
This research has
made use of the SIMBAD data base, operated at CDS, Strasbourg,
France. This publication makes use of data products from the Two
Micron All Sky Survey, which is a joint project of the University of Massachusetts and the Infrared Processing and Analysis
Center/California Institute of Technology, funded by the National
Aeronautics and Space Administration and the National Science
Foundation.

\bibliographystyle{mn2e}

\end{document}